\documentclass[aps,prl,twocolumn,superscriptaddress,nobibnotes]{revtex4-2}
\usepackage{amssymb}
\usepackage{booktabs}
\usepackage{physics}
\usepackage{graphicx}
\usepackage{amsmath}
\usepackage{amsthm}
\usepackage{amsfonts}
\usepackage[T1]{fontenc}
\usepackage{tikz}
\usepackage{array}
\usepackage{multirow}
\usepackage{color}
\usepackage{esint}
\usepackage{bm}
\definecolor{LinkColor}{rgb}{0.75,0.0,0.2}
\usepackage{bbm}
\usepackage[english]{babel}
\usepackage{hyperref}
\hypersetup{
	pdfauthor={good guys},
	pdftitle={good title},
	colorlinks=true,
	citecolor=LinkColor,
	linkcolor=LinkColor,
	urlcolor=LinkColor,
}

\begin{document}
\title{Post-Selection-Free Decoding of Measurement-Induced Area-Law Phases via Neural Networks}
	
\author{Hui Yu}
\affiliation{Beijing National Laboratory for Condensed Matter Physics and Institute of Physics, Chinese Academy of Sciences, Beijing 100910, China}

\author{Jiangping Hu}
\email{jphu@iphy.ac.cn}
\affiliation{Beijing National Laboratory for Condensed Matter Physics and Institute of Physics, Chinese Academy of Sciences, Beijing 100910, China}
\affiliation{New Cornerstone Science Laboratory, Institute of Physics, Chinese Academy of Sciences, Beijing 100910, China}

\author{Shi-Xin Zhang}
\email{shixinzhang@iphy.ac.cn}

\affiliation{Beijing National Laboratory for Condensed Matter Physics and Institute of Physics, Chinese Academy of Sciences, Beijing 100910, China}

\date{\today}
 
\begin{abstract}
\noindent 
Monitored quantum circuits host a rich variety of exotic non-equilibrium phases. Among the most representative examples are measurement-induced phase transitions between distinct area-law entangled states. However, because these transitions are characterized by specific entanglement quantities such as mutual information or topological entanglement entropy that are nonlinear functionals of the density matrix, their experimental observation requires multiple identical quantum trajectories via post-selection, which becomes exponentially unfeasible for large systems. Here, we leverage modern machine learning tools to address this challenge. We devise a neural network architecture combining a convolutional neural network with an attention mechanism, and use raw measurement outcomes directly as input to classify trivial, long-range entangled, and symmetry-protected topological phases. We show that the system's relaxation to a steady-state phase manifests as a sharp convergence in the classifier's accuracy, entirely bypassing the need for quantum state reconstruction. We systematically study the performance of our network as a function of sample size, input data, spatial and temporal constraints, and system size scalability. Our results demonstrate that this approach is robust and post-selection free, offering a practical pathway for experimentally probing measurement-induced phases. 
\end{abstract}

\maketitle
\noindent
The dynamics of quantum entanglement have emerged as a powerful paradigm for characterizing non-equilibrium many-body systems. Driven by the advent of noisy intermediate-scale quantum (NISQ) technologies \cite{preskill2018quantum}, random circuits \cite{fisher2023random,liu2025noisy} have become a versatile platform for studying these dynamics. A prominent example is the hybrid quantum circuit \cite{skinner2019measurement,li2018quantum,li2019measurement,gullans2020dynamical,chan2019unitary,bao2020theory,zabalo2020critical,choi2020quantum,jian2020measurement,turkeshi2020measurement,tang2020measurement,fan2021self,liu2024noise,liu2023universal,liu2024entanglement,qian2025protect}, where an entanglement phase transition arises from the competition between random unitary evolution and interspersed measurements. At high measurement rates, frequent projective operations collapse the wavefunction, driving the steady state from volume-law to area-law entanglement scaling.

While the transition from volume-law to area-law entanglement is well established, measurement induced transitions can also occur between distinct area-law phases, such as from a trivial phase to a long-range entangled (LR) \cite{sang2021measurement,lang2020entanglement,yu2025measurement,qian2024steering} or symmetry protected topological (SPT) phase \cite{lavasani2021measurement,morral2023detecting,yu2025gapless}. These transitions are driven solely by the competition between non-commuting projective measurements \cite{ippoliti2021entanglement}. Crucially, the associated entanglement transitions leave no signature in standard linear observables. Experimental verification requires preparing multiple identical copies of the same quantum state to measure quantities like entanglement entropy. Due to the intrinsic randomness of measurement outcomes, reproducing a specific trajectory requires an experimental overhead that scales exponentially with system size. This post-selection problem limits the experimental realizations of these transitions to small systems \cite{noel2022measurement,koh2023measurement,google2023measurement}.

Various strategies have been proposed to mitigate or bypass the post-selection problem. One prominent approach compares experimental outcomes with classical simulations via cross-entropy benchmarking \cite{li2023cross,kamakari2025experimental,tikhanovskaya2024universality}. Another exploits spacetime duality to map circuit dynamics to correlation functions \cite{ippoliti2022fractal,ippoliti2021postselection} in unitary circuits, avoiding the exponential overhead. Recent studies also suggest that critical properties of the transition can be extracted from early-time dynamics \cite{wang2024driven,wang2026relaxation}, which requires shallower circuit depths and reduces exponential post-selection cost. Further methods focus on the system's evolution and properties, such as employing scalable feedback protocols to actively steer the wavefunction \cite{o2024entanglement,roy2020measurement}, or interpreting the transition through the lens of quantum error correction capabilities \cite{gullans2020dynamical,gullans2020scalable,ippoliti2024learnability,dehghani2023neural}. Statistical techniques, like shadow tomography \cite{aaronson2018shadow,huang2020predicting,roser2025robust,hou2025machine}, have also been utilized to estimate lower and upper bounds on the entanglement entropy. However, most of these strategies are designed to distinguish between two specific phases, typically the volume-law and area-law regimes. Distinguishing multiple coexisting area-law phases driven by competing non-commuting measurements remains an open challenge.

Motivated by the recent success of machine learning (ML) algorithms in classifying quantum many-body phases \cite{carrasquilla2017machine,huang2022provably,carrasquilla2017machine}, we propose a neural network (NN) approach that combines a convolutional neural network (CNN) \cite{lawrence1997face,o2015introduction} with an attention mechanism \cite{vaswani2017attention,parikh2016decomposable,kim2025attention,kim2025learning} to classify three distinct area law phases with different entanglement structures: trivial, LR, and SPT. These phases arise from the competition among three non-commuting measurement gates $X$, $ZZ$, and $ZXZ$. First, we introduce the measurement-only circuit model and describe our physically motivated NN architecture. We then systematically examine how classification performance depends on key physical parameters: the number of sampled trajectories, the use of full versus restricted measurement data, the final evolution time, and the spatial extent of the input. We further demonstrate that a network trained on smaller systems generalizes well to larger system sizes. Finally, we evaluate the role of the attention mechanism by comparing our CNN+attention model with a CNN only baseline.

\begin{figure}
\centering
\includegraphics[width=0.5\textwidth, keepaspectratio]{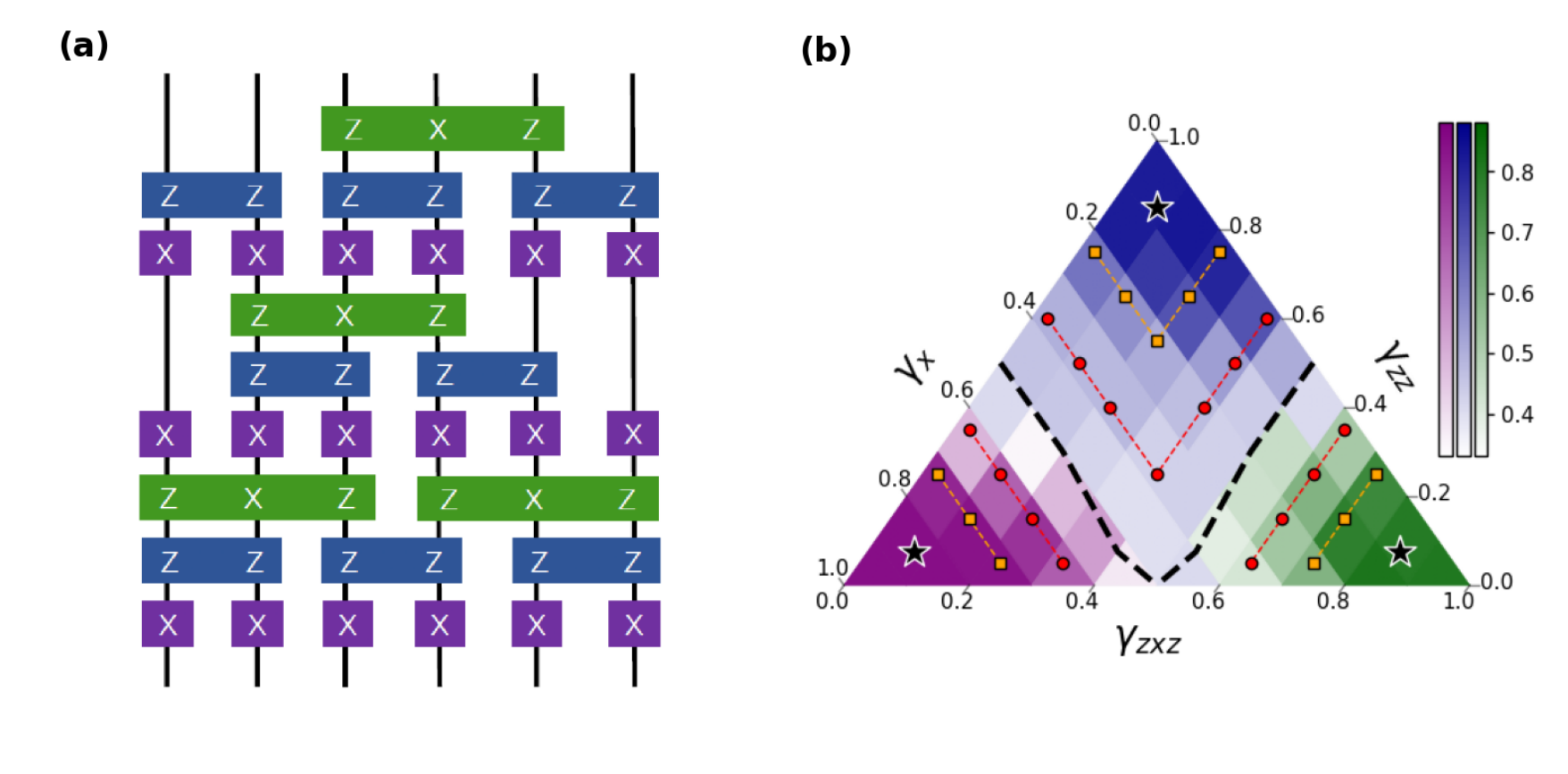}
\caption{Circuit architecture and phase diagram. (a) Schematic of the measurement-only brickwork circuit applied to $L=6$ qubits over three time steps. Each step features three sequential layers of measurements: single-qubit $X$ operators shown in purple, nearest-neighbor two-qubit $ZZ$ operator in blue, and three-qubit $ZXZ$ operators on adjacent triplets in green. (b) Ternary phase diagram defined on the simplex $\gamma_{X}+\gamma_{ZZ}+\gamma_{ZXZ}=1$. The purple, blue, and green shaded regions correspond to the trivial, long range entangled, and SPT phases, respectively, with dashed lines representing the critical boundaries between them. Stars at the three vertices mark the phase training points. Test parameter locations are indicated by red circles for outer points and orange squares for inner points. The color bar indicates the fraction of independent training runs that agree on the majority label.}
\label{fig:circuit}
\end{figure}

\vspace{\baselineskip}
\noindent {\fontsize{12}{14}\selectfont \textbf{Results}}

\noindent {\fontsize{10}{14}\selectfont \textbf{Model}}

\noindent
We investigate a measurement-only circuit defined on one-dimensional brickwork architecture of $L$ qubits with open boundary conditions, as depicted in Fig.~\ref{fig:circuit} (a). The system, prepared in an initial state $|\psi_{0}\rangle$, is evolved for $T$ time steps until a steady state is reached. Each time step comprises three sequential layers of generalized weak measurements (see \textbf{Methods}): $(1)$ single-qubit measurements of $X_{i}$, $(2)$ two-qubit measurements of $Z_{i}Z_{i+1}$ on nearest neighbor pairs $(i,i+1)$, and $(3)$ three-qubit measurements of $Z_{i}X_{i+1}Z_{i+2}$ on adjacent triplets $(i,i+1,i+2)$. The strength of each measurement type $Q \in \{X,ZZ,ZXZ\}$ is govern by a parameter $\gamma_{Q}$, subject to the constraint $\gamma_{X}+\gamma_{ZZ}+\gamma_{ZXZ}=1$. Crucially, we employ generalized weak measurements rather than standard projective ones ($\gamma_{Q}=1$). In the context of machine learning, the projective measurement introduce a shortcut, as the fraction of active measurement events directly reveals the measurement rate. The network can thus bypass learning entanglement structure and simply exploit the statistics of measurement occurrences to predict the phase.

As established in prior work \cite{yu2025gapless,yu2025measurement}, this circuit yields a phase diagram partitioned into three distinct regions, as shown in Fig.~\ref{fig:circuit} (b), which are distinguished by the entanglement entropy of their late-time steady states. In the limit $\gamma_{X} \rightarrow1$, single-qubit measurements dominate, collapsing the qubits into the $X$ basis and yielding a trivial, short-range entangled state. Conversely, when $\gamma_{ZZ}\rightarrow1$, two-qubit measurements build long range correlations across the chain, driving the steady state toward a macroscopic Greenberger-Horne-Zeilinger \cite{greenberger1989going} like state. Finally, as $\gamma_{ZXZ}\rightarrow1$, the system enters a SPT phase. This phase is protected by a global $\mathbb{Z}_{2}\times\mathbb{Z}_{2}$ symmetry generated by $\prod_{even}X_{i}$ and $\prod_{odd}X_{i}$ and is equivalent to the ground state of the cluster model \cite{son2011quantum,yu2023quantum,zheng2025emergence}. The critical boundaries separating these phases are identified through the finite-size scaling of the mutual information and the topological entanglement entropy \cite{kitaev2006topological,levin2006detecting}.

Unless otherwise specified, all numerical results presented below are obtained for a system of $L = 12$ qubits, and evolved for $T = 6L$ time steps. The system is initialized in the state $|\psi_{0}\rangle=(\frac{1}{\sqrt2})^{L}(|0\rangle+i|1\rangle)^{\bigotimes L}$.

\vspace{\baselineskip}
\noindent {\fontsize{10}{14}\selectfont \textbf{NN architecture}}

\begin{figure*}
\centering
\includegraphics[width=1\textwidth, keepaspectratio]{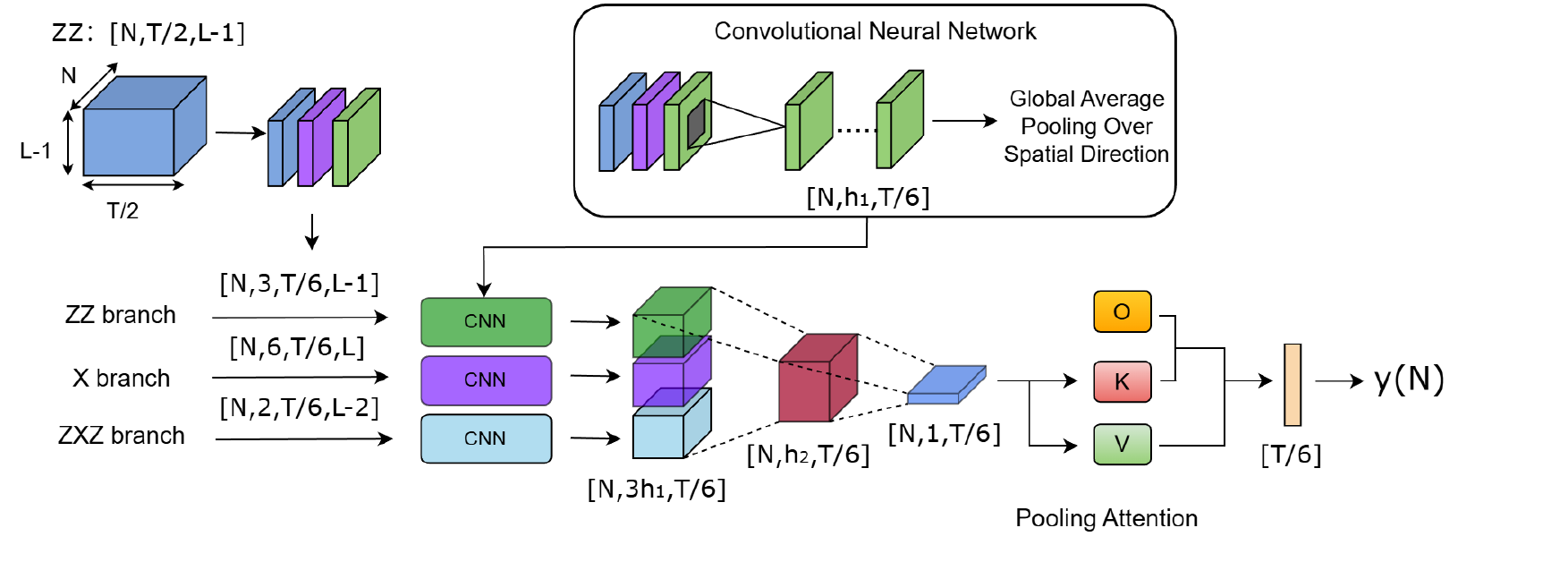}
\caption{Schematic of the neural network architecture. Raw $X$, $ZZ$, and $ZXZ$ measurement records for $N$ trajectories are reshaped to align with the circuit's brickwork periodicity. Three parallel CNN branches independently process these inputs, followed by spatial global average pooling to extract feature sequences. These features are concatenated, projected through a linear layer, and compressed via a temporal readout. Finally, a pooling attention module aggregates the $N$ trajectory representations into a single vector, which is mapped via a softmax layer to produce $y(N)$, a probability distribution over the trivial, LR, and SPT phases. Bracketed notation (e.g.,$[...]$) is provided at each intermediate stage to explicitly indicate the shape of the data tensors.}
\label{fig:NN}
\end{figure*}

\noindent
The NN architecture, illustrated in Fig.~\ref{fig:NN}, is designed to classify the physical phase of a given parameter point $(\gamma_{X}, \gamma_{ZZ}, \gamma_{ZXZ})$. The model utilizes a collection of $M$ measurement trajectories, which are divided into smaller, independent sets of size $N$. Each trajectory records the binary outcomes of all measurement gates applied during the circuit's evolution. The input to the NN consists of $N$ trajectories of measurement data associated with $X$, $ZZ$, and $ZXZ$. 
This input is processed through a CNN, several linear layers, and an attention mechanism (see \textbf{Methods} for details) to produce a three element output vector, $y(N)$, which represents a probability distribution over the three phase labels. To evaluate the full collection of $M$ trajectories, a final ensemble prediction, $y(M)$, is obtained by averaging the individual predictions $y(N)$ across all $M/N$ independent sets.

The network is trained on data from the three pure phase vertex points, corresponds to the parameter coordinates $(\gamma_{X},\gamma_{ZZ},\gamma_{ZXZ})=(0.85,0.075,0.075)$, $(0.075,0.85,0.075)$, and $(0.075,0.075,0.85)$. For both training and testing, we generate $M = 10000$ total trajectories per parameter point, partitioned into batches of $N = 25$. The network weights are optimized for $30$ epochs by minimizing the cross-entropy loss using the Adam optimizer with a learning rate of $2\times10^{-5}$. The cross-entropy loss function is defined as:  
\begin{equation}
\mathcal{L} = -\sum_{c=1}^{3} y_{c} \log \hat{y}_{c}
\end{equation}
where $c\in\{1,2,3\}$ indexes the three phase classes, $\hat{y}_{c}$ denotes the probability predicted by the NN for class $c$, and ${y}_{c}\in\{0,1\}$ is the binary indicator from the true label vector. A detailed analysis of the hyperparameter search is provided in the Supplemental Material (SM).

\vspace{\baselineskip}
\noindent {\fontsize{10}{14}\selectfont \textbf{Phase diagram}}

\noindent
To reconstruct the phase diagram, we sample test parameter points across the simplex and determine the phase of each point via majority vote from an ensemble of $30$ independently trained models. Every model computes a probability distribution $y(M)$ and casts a single vote for the phase with the highest predicted probability. The final phase assignment for a given parameter point is determined by the majority consensus across the ensemble, color-coded as purple (trivial), blue (LR), or green (SPT). Furthermore, the color intensity encodes the fraction of runs agreeing on the majority label, providing a visual measure of classification confidence. Comprehensive details regarding the specific sampling grid and the ensemble voting protocol are provided in the SM. The reconstructed phase diagram in Fig.~\ref{fig:circuit} (b) shows good overall agreement with the phase boundaries determined from prior works \cite{yu2025gapless,yu2025measurement}. Minor discrepancies arise predominantly near the critical lines, where the overlap of measurement statistics between adjacent phases is most pronounced and finite-size effects are the strongest.

\vspace{\baselineskip}
\noindent {\fontsize{10}{14}\selectfont \textbf{Impact of sample size and input channels}}
\begin{figure}
\centering
\includegraphics[width=0.5\textwidth, keepaspectratio]{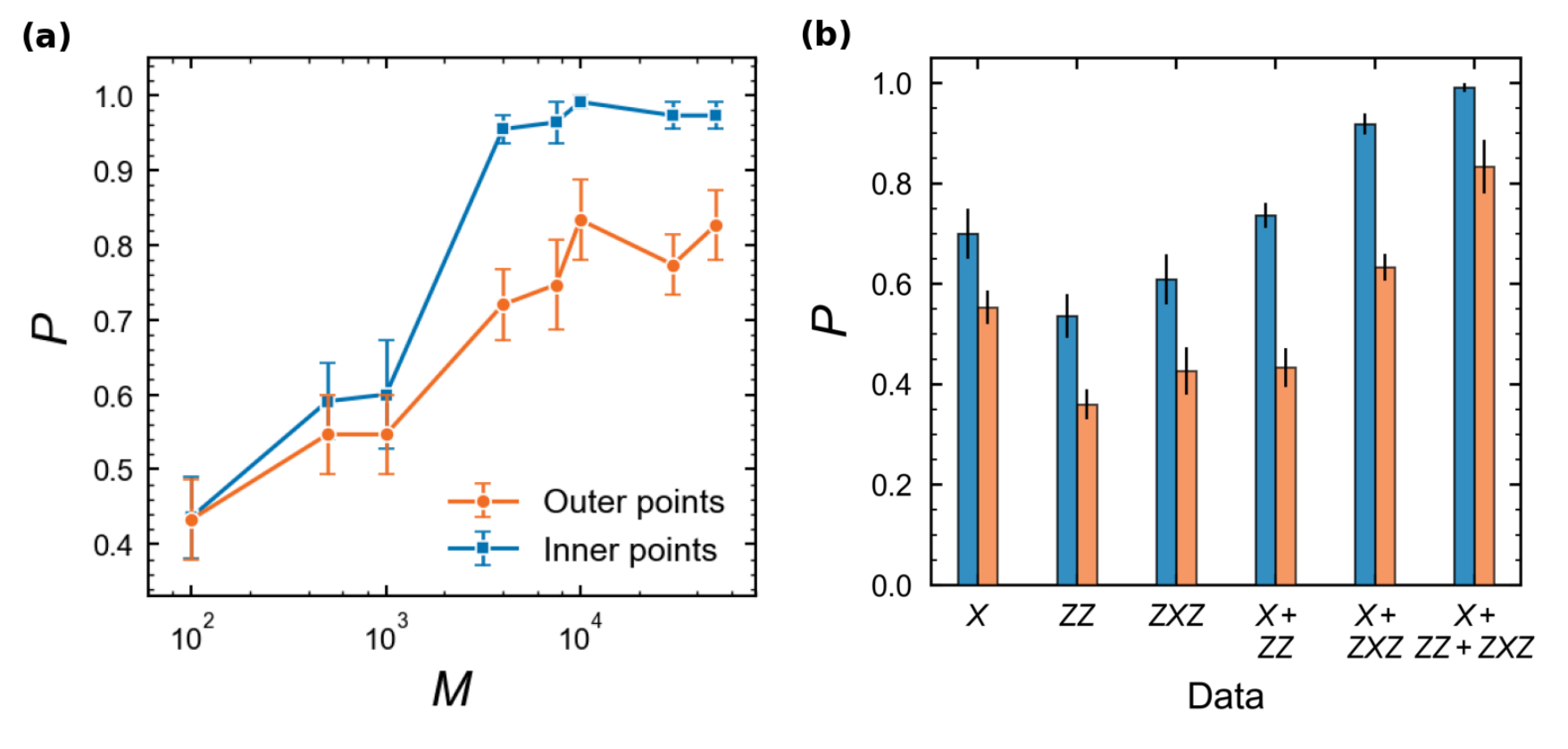}
\caption{Dependence of classification accuracy on trajectory count and input data type. (a) Classification accuracy $P$ as a function of the number of measurement trajectories $M$. Results are shown for $N=25$ across discrete values of $M\in\{100, 500, 1000, 4000, 7500, 10000, 30000, 50000\}$. (b) Classification accuracy $P$ across different combinations of input measurement channels. Blue and orange denote inner and outer test points, respectively. Error bars represent the standard error across 10 trained models from different random initializations.}
\label{fig:M_data_type}
\end{figure}

\noindent
We first examine the dependence of the classification accuracy $P$ on the number of sampled trajectories $M$ and the choice of input data. Defined as the fraction of correctly identified test points, $P$ is determined for each parameter point via a the majority vote of $10$ independently trained models. To ensure statistical reliability, we also average $P$ over $10$ independent repetitions of this voting protocol. A comprehensive description of the formal calculation of $P$ and the associated statistical averaging procedure is provided in the SM.

Figure~\ref{fig:M_data_type}(a) shows $P$ versus $M$ for inner points lie deep within each phase and outer points locate near the boundaries. For inner points, accuracy rises sharply and saturates near $P \approx 0.97$ around $M=3000$, as measurement records in the bulk of each phase are highly distinctive. Conversely, outer points require more trajectories to achieve reliable classification, reaching $P \approx 0.83$ at $M=10000$. This difference highlights the intrinsically harder nature of near-boundary classification, where the measurement statistics of adjacent phases become increasingly similar. Furthermore, the consistently larger error bars for the outer points compared to the inner points reflect the greater classification difficulty near the phase boundaries.

Fig.~\ref{fig:M_data_type}(b) examines the impact of the input data channels. Using a single measurement channel in isolation yields poor accuracy below $0.6$ for outer points, showing that no individual channel can reliably discriminate all three phases. Combining two channels improves performance, while using all three channels achieves the highest accuracy, reaching near unity for inner points and $P$ over $0.8$ for outer points. This proves that the measurements carry complementary information. Specifically, the $X$, $ZZ$, and $ZXZ$ channels are most sensitive to the trivial, LR, and SPT phases. This strong synergy directly justifies our multi-branch network design.

\vspace{\baselineskip}
\noindent {\fontsize{10}{14}\selectfont \textbf{Impact of circuit depth and spatial subsystem size}}

\begin{figure}
\begin{center}
\includegraphics[width=0.5\textwidth, keepaspectratio]{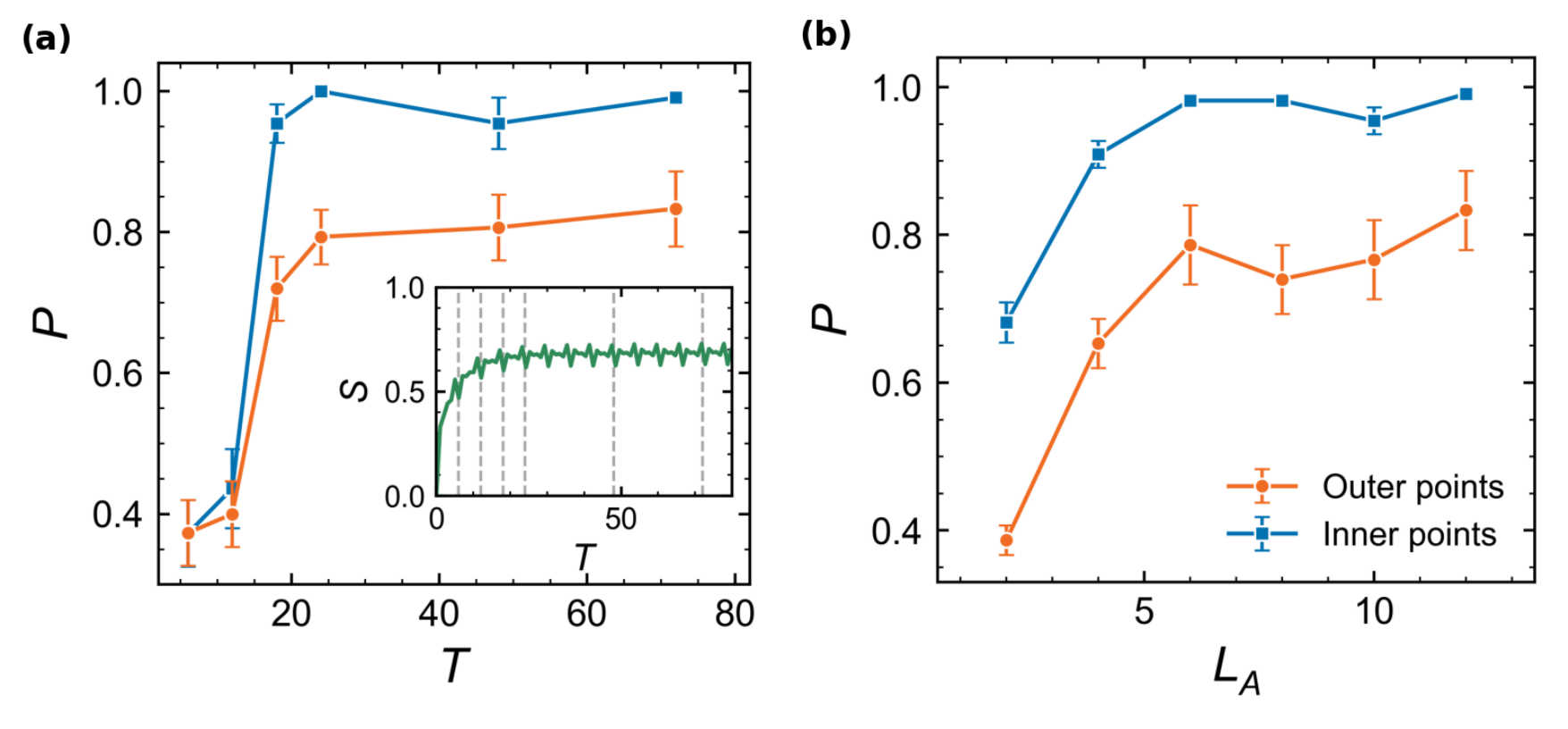}
\caption{Dependence of classification accuracy on evolution time and spatial subsystem size. (a) Classification accuracy $P$ as a function of the total evolution time $T$, evaluated at discrete times $T \in \{0.5L, L, 1.5L, 2L, 4L, 6L\}$ with $L=12$. The inset shows the half chain entanglement entropy $S$ versus $T$ at $\gamma_{X}=\gamma_{ZXZ}=0.3$. Dashed vertical lines indicate the specific evaluation times plotted in the main panel. (b) Classification accuracy $P$ versus spatial subsystem size $L_{A} \in \{2, 4, 6, 8, 10, 12\}$. In both panels, blue and orange markers correspond to inner and outer test points, respectively. Error bars denote the standard error of the mean across 10 independent trained models from different random initializations.}
\label{fig:TandL}
\end{center}
\end{figure}

\noindent
The temporal dependence of the accuracy $P$ is illustrated in Figure~\ref{fig:TandL}(a). At early times $T \lesssim 10$, the circuit has not yet relaxed to its steady state. Consequently, the measurement records lack distinct phase signatures, resulting in near-random network performance. As the evolution progresses, the accuracy rises sharply, saturating near perfect classification around $T \approx 24$ for inner points. For outer points, this saturation occurs more gradually, eventually plateauing near $P \approx 0.8$. Crucially, this performance plateau coincides perfectly with the saturation of the half-chain entanglement entropy $S$ shown in the inset, confirming that the system relaxes to its physical steady state on this exact timescale. This correspondence between the convergence of entanglement and the onset of reliable classification provides a physical justification for our chosen circuit depth.

Turning to the spatial constraints, we evaluate the network's performance when access is limited to a contiguous subsystem of size $L_{A}$, as detailed in Fig.~\ref{fig:TandL}(b). The subsystem is defined as the central segment of the chain, extending from the $(L-L_{A})/2$-th qubit to the $(L+L_{A}-2)/2$-th qubit. Note that for $L_{A}=2$, the $ZXZ$ channel is absent since it requires at least three contiguous qubits, and the network receives only $X$ and $ZZ$ measurement records. Remarkably, inner points already achieve
$P \approx 0.7$ even in this minimal two-qubit setting, confirming that essential phase distinguishing features are locally encoded within the measurement records. As $L_{A}$ grows, the accuracy rises sharply, saturating near unity for inner points around $L_{A}=6$. By contrast, outer points exhibit a more gradual improvement, scaling from $P \approx 0.4$ at $L_{A}=2$ up to $P \approx 0.8$ for the full chain. This aligns perfectly with the expectation that accurately resolving near-boundary states necessitates access to a larger portion of the system. Nonetheless, the substantial information content present at small scales demonstrates that partial measurement records drawn from a restricted spatial region are highly effective for robust phase identification.

\vspace{\baselineskip}
\noindent {\fontsize{10}{14}\selectfont \textbf{Scalability and role of the attention mechanism}}
\begin{figure}
\centering
\includegraphics[width=0.495\textwidth, keepaspectratio]{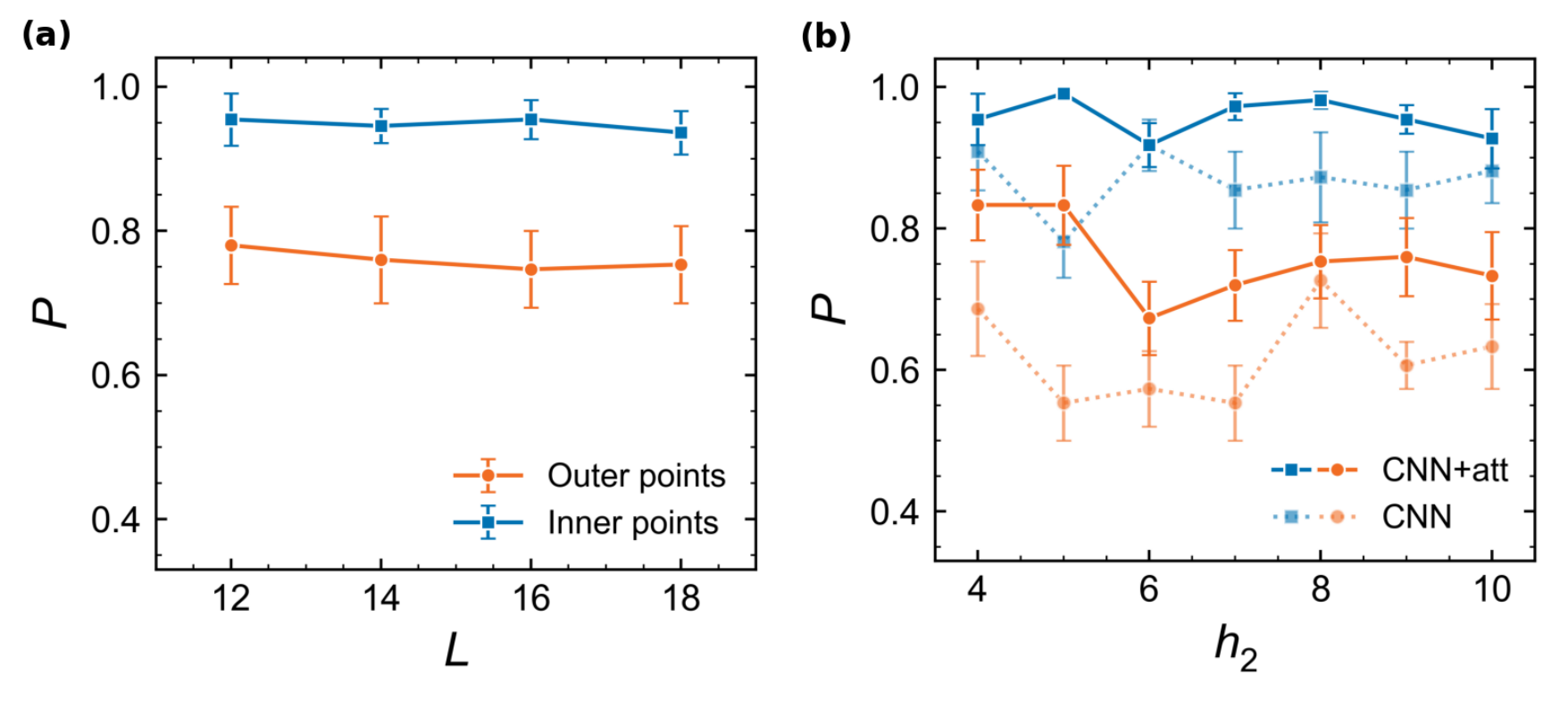}
\caption{Transfer learning scalability and impact of the attention mechanism. (a) Classification accuracy $P$ versus system size $L$ for a network trained on $L = 12$ and $T=36$ and tested on larger systems without retraining ($h_{1}=8$, $h_{2}=5$). (b) Classification accuracy $P$ for the model with (dark) and without (light) the attention mechanism with $h_{1}=8$. In both panels, blue and orange markers correspond to inner and outer test points, respectively. Error bars denote the standard error of the mean across 10 independent trained models.}
\label{fig:transfer}
\end{figure}

\noindent
A key practical requirement for any ML-based phase classifier is the ability to generalize across system sizes without retraining. Figure~\ref{fig:transfer}(a) shows $P$ as a function of system size $L$ when the network is trained exclusively on $L = 12$ data and evaluated on $L=14$, $16$, and $18$ qubits. For evaluations on larger systems, the measurement records are drawn from a central $12$-qubit subsystem to maintain a consistent input dimension for the convolutional layers. The inner point accuracy remains consistently high near $0.95$ across all tested sizes with minor fluctuations. The outer point accuracy stays above $0.75$ and exhibits a minor degradation at $L=14$ before stabilizing at $L=16$ and $L=18$. This size transferability indicates that the learned representations capture universal, size-independent features of the three phases rather than artifacts specific to $L = 12$. 

To quantify the contribution of the attention mechanism, we compare the full CNN+attention model against a CNN-only baseline across a range of parameters in Fig.~\ref{fig:transfer}(b). With $h_{1}$ fixed at $8$, the CNN+attention model consistently outperforms the baseline across all values of $h_{2}$ for both inner and outer test points. This performance gap is most significant for outer points near the phase boundaries. Here, the CNN-only model achieves an average $P$ of $0.6$, whereas the CNN+attention model reaches $0.75$. Furthermore, we evaluate a Multilayer Perceptron (MLP) baseline that processes trajectories individually without any collective aggregation. As detailed in the SM, the MLP exhibits the poorest performance among all architectures, with accuracy dropping to approximately $0.5$
for outer test points. These results suggest that the attention mechanism effectively classifies phases by capturing correlations between trajectories. Furthermore, the higher variance observed in the CNN-only model for inner points suggests that, without the temporal aggregation of the attention module, the network remains significantly more sensitive to random training initialization.

\vspace{\baselineskip}
\noindent {\fontsize{12}{14}\selectfont \textbf{Discussion}}

\noindent 
We demonstrate that a CNN architecture, augmented by an attention mechanism, accurately classifies the three distinct area-law phases of a measurement-only quantum circuit using raw measurement outcomes. By operating directly on classical trajectory records, our approach is inherently post-selection free, bypassing the need for quantum state reconstruction or the computation of complex nonlinear observables. A systematic analysis reveals that integrating all measurement channels and achieving the steady state are critical for high classification accuracy. Furthermore, we show that the network generalizes its learned universal features to larger system sizes without additional training, while the attention mechanism provides a robust performance advantage over baseline models near phase boundaries.

A natural extension of this work is to apply this framework to circuits with more complex measurement structures, including a broader range of topological phases beyond $\mathbb{Z}_{2}\times\mathbb{Z}_{2}$ SPT \cite{chen2013symmetry,tsui2017phase} class. Additionally, investigating the interpretability of the network's learned features will help elucidate the specific spatiotemporal patterns in the measurement records that drive phase distinction. More broadly, this study advances the intersection of quantum information and machine learning, offering a promising pathway for the design of efficient classical post-processing protocols for experimental verification of measurement-induced phases.

\vspace{\baselineskip}
\noindent {\fontsize{12}{14}\selectfont \textbf{Methods}}

\noindent {\fontsize{10}{14}\selectfont \textbf{Generalized weak measurements}}

\noindent Mathematically, the weak measurement \cite{nielsen2010quantum} for an observable $Q$ (with eigenvalue $\pm1$) is implemented as a quantum channel defined by Kraus operators $\{ M_{1}^{Q},M_{2}^{Q}\}$. Let $\Pi_{\pm}^{Q} = \frac{I \pm Q}{2}$ denote the projectors onto the $\pm1$ eigenspaces of $Q$. We construct the Kraus operators as:
\begin{equation}
\begin{split}
M_{1}^{Q} &= \cos{(\theta_{Q})}\Pi_{+}^{Q}+\sin{(\theta_{Q})}\Pi_{-}^{Q} \\
M_{2}^{Q} &= \sin{(\theta_{Q})}\Pi_{+}^{Q}+\cos{(\theta_{Q})}\Pi_{-}^{Q}
\end{split}
\end{equation}
where the mixing angle is given by $\theta_{Q}=(1-\gamma_{Q})\frac{\pi}{4}$. These operators natively satisfy the completeness relation $\sum_{k=1,2}M_{k}^{Q\dagger}M_{k}^{Q}=I$. This parameterization smoothly interpolates between two extremes. When $\gamma_{Q}=0$ $(\theta_{Q}=\pi/4)$, the Kraus operators become $M_{1}^{Q}=M_{2}^{Q}=\frac{1}{\sqrt2}I$, corresponding to the absence of measurement. Conversely, when $\gamma_{Q}=1$ $(\theta_{Q}=0)$, the operators reduce to $M_{1}^{Q}=\Pi_{+}^{Q}$ and $M_{2}^{Q}=\Pi_{-}^{Q}$, thereby recovering standard projective measurements. 

After each measurement, the wavefunction is explicitly normalized  to the proper post-measurement state: 
\begin{equation}
|\psi_{k}\rangle = \frac{M_{k}^{Q}|\psi\rangle}{\sqrt{p_{k}}}
\end{equation}
where the probability of outcome $k\in\{1,2\}$ is
$p_{k}=\langle\psi|M_{k}^{Q\dagger}M_{k}^{Q}|\psi\rangle$. Circuit simulations are executed using the {\sf TensorCircuit-NG} \cite{zhang2023tensorcircuit,zhang2026tensorcircuit} package.

\vspace{\baselineskip}
\noindent {\fontsize{10}{14}\selectfont \textbf{Neural network architecture and pipeline}}

\noindent
Here we describe the complete pipeline for our neural network, detailing the data flow from input to final prediction. We begin with a system of $L$ qubits evolved for $T$ time steps, generating a total of $M$ measurement trajectories. These trajectories are then partitioned into $M/N$ sets, each containing $N$ trajectories. To mitigate overfitting, the full set of $M$ trajectories is randomly shuffled and repartitioned at the start of each epoch.

The three measurement types produce separate records: the $X$ channel records outcomes on $L$ qubits at $T$ steps, yielding a tensor shape of $[N,T,L]$; the $ZZ$ channel records outcomes on $L-1$ nearest-neighbor bonds at $T/2$ steps, producing a shape of $[N,T/2,L-1]$; and the $ZXZ$ channel records outcomes on $L-2$ triplets at $T/3$ steps, resulting in a shape of $[N,T/3,L-2]$. Before being passed to the CNN, each tensor is reshaped into a 4-dimensional format. Specifically, the $X$, $ZZ$, and $ZXZ$ tensors are reshaped to $[N,6,T/6,L]$, $[N,3,T/6,L-1]$, and $[N,2,T/6,L-2]$, respectively.

Following this preprocessing, three independent CNN branches process the $X$, $ZZ$, and $ZXZ$ channels in parallel. Each branch utilizes a two-dimensional convolutional window featuring a $3 \times 3$ kernel, a stride of $1$, and a padding of $1$ along the temporal and spatial axes. This operation is immediately followed by batch normalization (BN) and a ReLU activation. Let $x^{(b)}$ be the input tensor for branch $b \in \{X, ZZ, ZXZ\}$. The feature map $F^{(b)}$ is given by:
\begin{equation}
F^{(b)} = \text{ReLU}\left( \text{BN}\left( x^{(b)} \ast W^{(b)} \right) \right),
\end{equation}
where $\ast$ denotes the 2D convolution operation and $W^{(b)}$ contains the learnable convolutional weights. This operation is followed by global average pooling over the spatial dimension $L_{b}$, producing a feature tensor $H^{(b)}$ of shape $[N, h_{1},T/6]$, where $h_{1}$ denotes the number of convolutional output channels:
\begin{equation}
H^{(b)}_{n,c,\tau} = \frac{1}{L_b} \sum_{l=1}^{L_b} F^{(b)}_{n, c, \tau, l}.
\end{equation}
where $n$, $c$, and $\tau$ are indices corresponding to the set size $N$, the feature channels $h_{1}$, and the temporal steps $T/6$, respectively. 

Next, the outputs from the three branches are then concatenated along the feature axis to form a composite tensor $H^{cat}$ of shape $[N, 3h_{1}, T/6]$. A subsequent module projects this tensor down to a hidden representation $H^{f}$ of a shape $[N,h_{2},T/6]$, with $h_{2}$ serving as an additional architectural hyperparameter. This module applies a linear transformation followed by a ReLU activation, layer normalization (LN), and dropout regularization with a probability of $p=0.2$:   
\begin{align}
H^{f} &= \text{LN}\left( \text{ReLU}\left( H^{cat} W_{cat} + b_{cat} \right) \right)
\end{align}
where $W_{cat} \in \mathbb{R}^{3h_1 \times h_2}$ is the learnable weight matrix and $b_{cat} \in \mathbb{R}^{h_2}$ is the corresponding learnable bias vector. Finally, a temporal readout layer reduces the feature dimension at each time step to a scalar. This operation results in a one-dimensional sequence $Z$ of shape $[N, T/6]$:
\begin{equation}
Z = H^{f} W_{f} + b_{f},
\end{equation}
where $W_{f} \in \mathbb{R}^{h_2 \times 1}$ is the learnable weight matrix and $b_{f} \in \mathbb{R}$ is the learnable scalar bias.

To aggregate the $N$ trajectory representations into a single phase prediction, we employ a pooling attention module. A learnable query vector $O \in \mathbb{R}^{1 \times T/6}$ computes scaled dot-product attention scores against keys $K=Z W_K$ and values $V=Z W_V$, where $W_K,W_V\in \mathbb{R}^{T/6 \times T/6}$ are trainable projection matrices. This collapses the trajectory dimension $N$, producing an aggregated vector $q \in \mathbb{R}^{1 \times T/6}$ via a residual connection: 
\begin{equation}
q = O + \text{Softmax}\left( \frac{O K^{\top}}{\sqrt{T/6}} \right) V.
\end{equation}
The aggregated vector $q$ is then processed through a module consisting of layer normalization and a single linear layer followed by ReLU with a residual connection, yielding a refined vector $q_{out}$:
\begin{equation}
q_{out} = \text{LN}\left( \text{LN}(q) + \text{ReLU}\left(\text{LN}(q) W_{q} + b_{q}\right) \right).
\end{equation}
Finally, $q_{out}$ is fed into a fully connected linear layer whose outputs are converted to a probability distribution $y(N)$ over the three phase labels via a softmax:
\begin{equation}
y(N) = \text{Softmax}\left( q_{out} W_{out} + b_{out} \right).
\end{equation}
The final prediction $y(M)$ for all $M$ trajectories is obtained by averaging the predictions across all $M/N$ sets.  
\begin{equation}
y(M) = \frac{N}{M}\sum_{i=1}^{M/N}y(N_{i})
\end{equation}

\vspace{\baselineskip}
\noindent {\fontsize{12}{14}\selectfont \textbf{Data availability}}

\noindent Numerical data for this manuscript are publicly accessible on Zenodo at \url{https://zenodo.org/records/19412221}.

\vspace{\baselineskip}
\noindent {\fontsize{12}{14}\selectfont \textbf{Code availability}}

\noindent All code used in this study is available on Zenodo at \url{https://zenodo.org/records/19412221}.

\vspace{\baselineskip}
\noindent {\fontsize{12}{14}\selectfont \textbf{References}}

\bibliography{reference}

\newpage 

\vspace{\baselineskip}
\noindent {\fontsize{12}{14}\selectfont \textbf{Acknowledgements}}

\noindent We acknowledge the support by the Ministry of Science and Technology  (Grant No. 2022YFA1403900), the National Natural Science Foundation of China (Grant No. NSFC-12494594) and the New Cornerstone Science Foundation. HY is also supported by the International Young Scientist Fellowship of Institute of Physics Chinese Academy of Sciences (No.202407). SXZ acknowledge the support
from Innovation Program for Quantum Science and
Technology (2024ZD0301700) and the National Natural Science Foundation of China (No. 12574546).

\vspace{\baselineskip}
\noindent {\fontsize{12}{14}\selectfont \textbf{Author contributions}}

\noindent H.Y., J.P.H., and S.-X.Z. conceived the project and designed the research. H.Y. performed the numerical calculations and data analysis under the supervision of S.-X.Z. and J.P.H. All authors wrote the manuscript and discussed the results.

\vspace{\baselineskip}
\noindent {\fontsize{12}{14}\selectfont \textbf{Competing interests}}

\noindent The authors declare no competing interests.

\end{document}

\renewcommand{\thefigure}{S\arabic{figure}}
\setcounter{figure}{0}
\renewcommand{\theequation}{S\arabic{equation}}
\setcounter{equation}{0}
\renewcommand{\thesection}{\Roman{section}}
\setcounter{section}{0}
\setcounter{secnumdepth}{4}

\title{Supplemental Material for ``Post-Selection-Free Decoding of Measurement-Induced Area-Law Phases via Neural Networks''}

\author{Hui Yu}
\affiliation{Beijing National Laboratory for Condensed Matter Physics and Institute of Physics, Chinese Academy of Sciences, Beijing 100910, China}

\author{Jiangping Hu}
\email{jphu@iphy.ac.cn}
\affiliation{Beijing National Laboratory for Condensed Matter Physics and Institute of Physics, Chinese Academy of Sciences, Beijing 100910, China}
\affiliation{New Cornerstone Science Laboratory, Institute of Physics, Chinese Academy of Sciences, Beijing 100910, China}

\author{Shi-Xin Zhang}
\email{shixinzhang@iphy.ac.cn}

\affiliation{Beijing National Laboratory for Condensed Matter Physics and Institute of Physics, Chinese Academy of Sciences, Beijing 100910, China}
\maketitle

\tableofcontents
\section{Sensitivity analysis of architectural hyperparameters}

The proposed neural network architecture is characterized by two primary hyperparameters:  $h_{1}$, the number of output channels in each convolutional neural network (CNN) branch, and $h_{2}$, the hidden dimension of the subsequent fusion layer. To evaluate their individual contributions to classification performance, we conduct two independent parameter scans by varying one hyperparameter while keeping the other constant.

\begin{figure}[ht]
\centering
\includegraphics[width=0.75\textwidth, keepaspectratio]{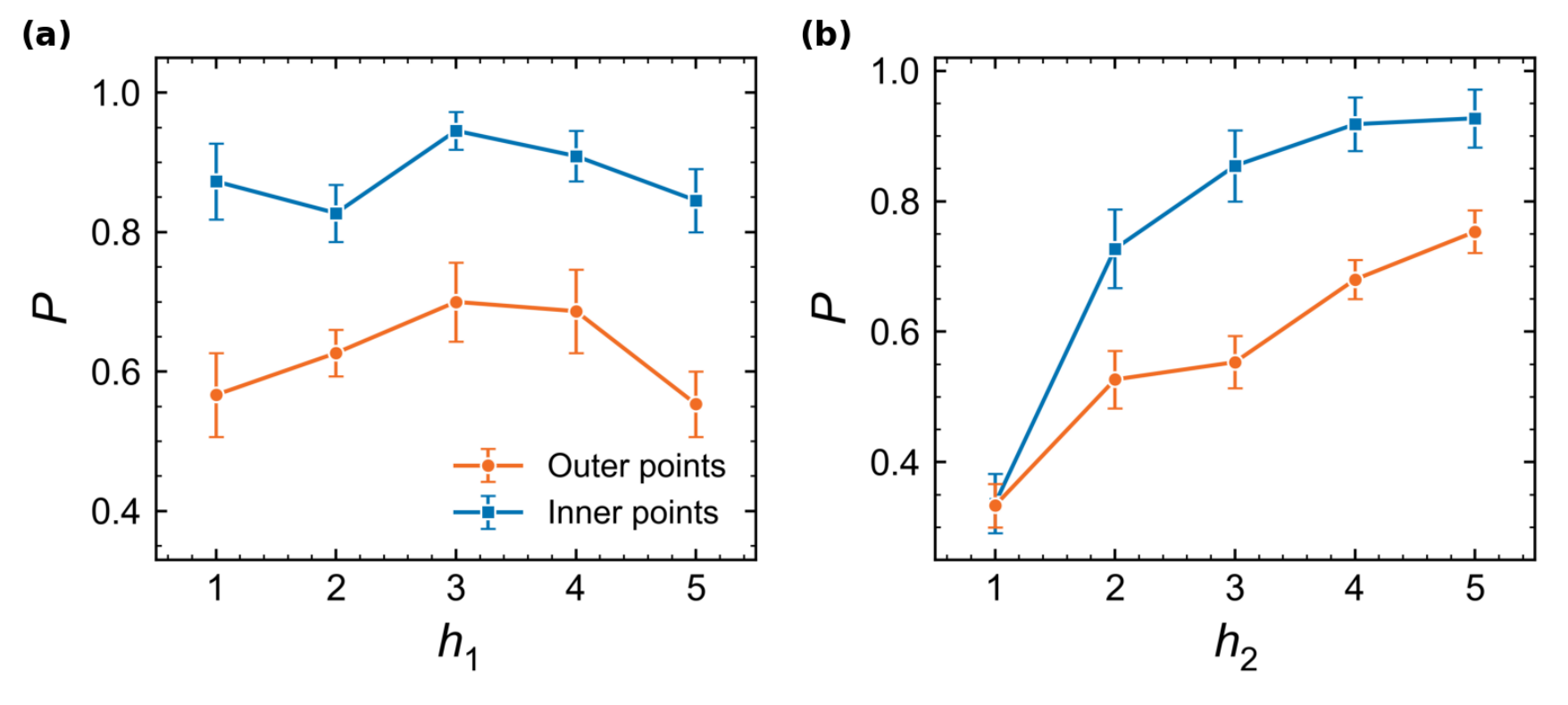}
\caption{Classification accuracy $P$ as a function of architectural hyperparameters. (a) $P$ versus the number of CNN branch output channels $h_{1}$, with the fusion hidden dimension fixed at $h_{2}=2$. (b) $P$ versus the fusion hidden dimension $h_{2}$ with the branch channels fixed at $h_{1}=1$. Blue and orange markers correspond to inner and outer test points, respectively. Error bars represent the standard error across 10 independent repetitions.}
\label{fig:parameter}
\end{figure}

Figure~\ref{fig:parameter}(a) illustrates the classification accuracy $P$ as a function of $h_{1}$ with $h_{2}$ fixed at $2$. The performance exhibits  minor fluctuation across both inner and outer test points as $h_{1}$ increases, suggesting that the network is relatively insensitive to the number of CNN output channels. In contrast, Figure~\ref{fig:parameter}(b) shows a distinct trend during the scan of $h_{2}$ with $h_{1}$ fixed at $1$. Accuracy scales positively with the hidden dimension. As $h_{2}$ increases from $1$ to $5$, $P$ rises from $0.33$ (equivalent to random guessing) to $0.9$ for inner points and $0.75$ for outer points. This improvement suggests that a larger $h_{2}$ enables the fusion layer to construct a more expressive joint representation across the three measurement channels. For inner points, we observe that moderate values of both $h_{1}$ and $h_{2}$ can achieve accuracies exceeding $0.9$. These preliminary results motivate a comprehensive grid search over $(h_{1},h_{2})$ parameter space to identify the optimal configuration, as detailed in the following section.

\section{Hyperparameter optimization via grid search}

\begin{figure}[ht]
\centering
\includegraphics[width=0.85\textwidth, keepaspectratio]{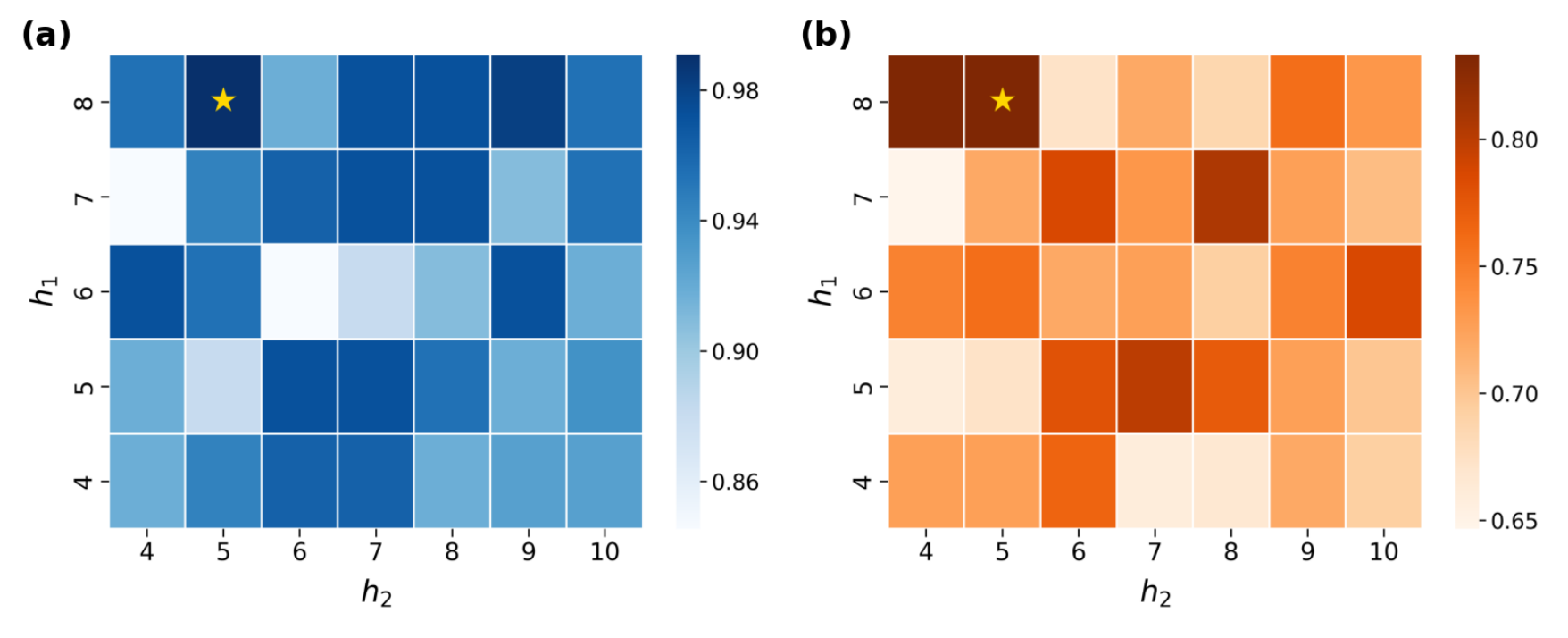}
\caption{Classification accuracy $P$ across the hyperparameter space $(h_{1},h_{2})$. Heat maps show the accuracy for (a) inner and (b) outer test points. The color intensity represents the mean accuracy $P$ averaged over 10 independent runs. The optimal configuration, $h_{1}=8$ and $h_{2}=5$, is indicated by a yellow star.}
\label{fig:optimal}
\end{figure}

To identify the optimal hyperparameter configuration, we conduct a systematic grid search across the two-dimensional $(h_{1}, h_{2})$ parameter space. Both hyperparameters are varied in moderate ranges: $h_{1}$ from $4$ to $8$ and $h_{2}$ from $4$ to $10$. For each coordinate in this space, we evaluate the classification accuracy $P$ for both inner and outer test points. As illustrated in Fig.~\ref{fig:optimal}(a), the accuracy for inner test points remains consistently high, with $P \geq0.85$ across the nearly sampled space. However, accuracy for outer test points is lower, with the lowest accuracy reaching approximately $0.65$ (Fig.~\ref{fig:optimal}(b)). This discrepancy is expected, as classification is inherently more challenging for points located near phase boundaries. The best overall performance is achieved at $h_{1}=8$ and $h_{2}=5$, yielding an accuracy of $P=0.99$ for inner points and $P=0.83$ for outer points. Based on these results, we adopted this configuration as the fixed architecture for all results reported in the main text. 

\section{Sensitivity to training and inference batch sizes}

\begin{figure}[ht]
\centering
\includegraphics[width=0.75\textwidth, keepaspectratio]{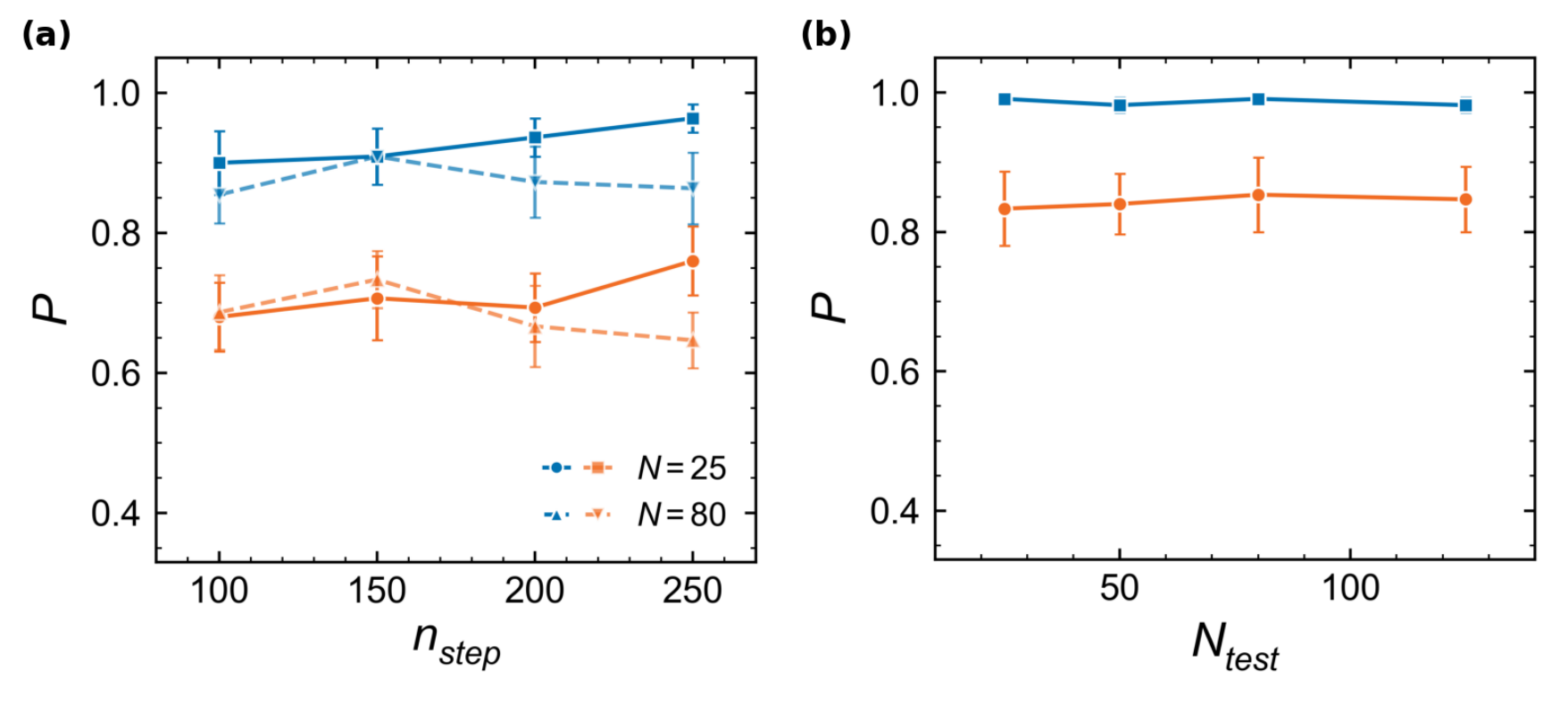}
\caption{Dependence of classification accuracy $P$ on batch size. (a) $P$ versus the number of stochastic resampling steps $n_{\mathrm{step}}$ for two training and testing batch sizes, $N=25$ (solid lines) and $N=80$ (dashed lines). In each step, $N$ trajectories are drawn with replacement from a pool of $10000$. (b) $P$ versus test batch size $N_{\mathrm{test}}$ for a fixed model trained with $N=25$, $h_{1}=8$, $h_{2}=5$. Blue and orange markers correspond to inner and outer test points, respectively. Error bars represent the standard error across $10$ independent runs.}
\label{fig:batch size}
\end{figure}

In this section, we investigate how the classification accuracy $P$ depends on the trajectory batch size $N$, using the optimal hyperparameters $(h_{1}=8,h_{2}=5)$. We present two complementary studies, illustrated in Fig.~\ref{fig:batch size}(a) and (b).

Fig.~\ref{fig:batch size}(a) examines the case where the training and testing batch sizes are identical. Here, rather than partitioning the $M=10000$ trajectories into fixed, independent sets, we employ a stochastic resampling strategy with replacement. For each of the $n_{\mathrm{step}}$ training steps, a fresh batch of $N$ trajectories is sampled uniformly at random from the total pool of $10000$. Consequently, individual trajectories may be sampled multiple times or not at all within a given epoch. The final prediction is obtained by averaging the network output $y(N)$ over $n_{step}$ iterations, resulting in a total of $N \times n_{\mathrm{step}}$ trajectory evaluations per epoch. We compare two batch sizes, $N=25$ and $N=80$, across a range of $n_{\mathrm{step}} \in \{100, 150, 200, 250\}$. For both cases, $P$ initially increases with $n_{\mathrm{step}}$ up to $150$. For $N=25$,  increasing $n_{step}$ further continues to improve performance because the total number of evaluations is still below the total pool of $10000$, allowing the network to better generalize from the available data. In contrast, for $N=80$, performance degrades as $n_{step}$ increases toward $250$. In this regime, the total number of evaluations significantly exceeds the $10000$ unique trajectories available. This forces the network to repeatedly process the same trajectories, leading to overfitting and a subsequent drop in test accuracy.

In Fig.~\ref{fig:batch size}(b), we assess the model’s robustness during inference. We fix the weights of a model trained at $N=25$ and vary only the test batch size, $N_{\mathrm{test}}$, from $25$ to $200$. The accuracy $P$ remains essentially stable across this entire range for both inner and outer test points. This stability demonstrates that the network's predictions are robust to changes in the inference scale.

\section{Phase identification and ensemble voting protocol}
To reconstruct the ternary phase diagram presented in the main text, we employ a systematic sampling and voting procedure. The process for determining the physical phase for any given test parameter point $(\gamma_{X},\gamma_{ZZ},\gamma_{ZXZ})$ is as follows. We first sample 
test points on the simplex defined by the constraint $\gamma_{X}+\gamma_{ZZ}+\gamma_{ZXZ}=1$. The grid is generated by varying $\gamma_{ZXZ}$ from $0.025$ to $0.925$ in increments of $0.1$, and for each value of $\gamma_{ZXZ}$, the parameter $\gamma_{X}$ is swept from $0.025$ to $0.95-\gamma_{ZXZ}$ in steps of $0.1$. For every sampled point, $M=10000$ independent measurement trajectories are generated and partitioned into $J=400$ sets, each containing $N=25$ trajectories.

Next, we utilize an ensemble of $30$ independent trained neural network models. Here, a trained model refers to a network instance whose weights and biases have been optimized using the Adam optimizer on labeled training data corresponding to the three pure phase vertex points. The models are independent in the sense that each is initialized with a unique random seed to mitigate individual model bias. For a specific parameter point, each model $i$ ($i=1,...,30$) processes all $J$ sets of measurement trajectories to compute an ensemble-averaged probability distribution over the $M$ trajectories:
\begin{equation}
y_{i}(M) = \frac{1}{J}\sum_{j=1}^{J}y_{i}(N_{j})
\end{equation}
where $y_{i}(N_{j})$ is the three-element probability vector predicted by model $i$ for the $j-$th set of $N$ trajectories. Each model then casts a single vote $v_{i}$ for the phase that possesses the highest probability:
\begin{equation}
v_i = \underset{c \in \{1,2,3\}}{\arg\max} \left[ y_i(M) \right]_c
\end{equation}
where $c$ indexes the trivial ($1$), long range entangled ($2$), and symmetry protected topological ($3$) phases. The final phase assignment for the parameter point is determined by the majority consensus of these $30$ votes.

\section{Accuracy metrics $P$ and ensemble averaging protocol}
The classification accuracy $P$ is the primary metric used to evaluate the network's performance across different physical regimes. It is defined as the fraction of correctly classified test points within a specific set $S$:
\begin{equation}
P = \frac{1}{|\mathcal{S}|} \sum_{s \in \mathcal{S}} \mathbf{1}\!\left[\hat{y}(s) = y(s)\right], 
\label{P}
\end{equation}
where $y(s)$ is the ground truth label for parameter point $s$, and $\mathbf{1}[\cdot]$ is the indicator function, which yields $1$ if the prediction matches the ground truth and $0$ otherwise. For our analysis, the test set $\mathcal{S}$ is categorized into two groups: inner points, located deep within the phase bulk, and outer points, situated in close proximity to the phase boundaries.

The predicted label $\hat{y}(s)$ is obtained following a procedure similar to the phase determination, but utilizes a smaller ensemble to evaluate general performance of the architecture. Specifically, $\hat{y}(s)$ is determined by the majority consensus of $10$ independently trained models. For each parameter point $s$, every model $i$ ($i=1,...,10$) computes its average probability distribution $y_{i}(M)$ (with $M=10000$ and $N=25$) and casts a vote $v_{i}$ for the phase with the highest probability:
\begin{equation}
v_i = \underset{c \in \{1,2,3\}}{\arg\max} \left[ y_i(M) \right]_c
\end{equation}
The ensemble prediction $\hat{y}(s)$ is then defined as the majority of these $10$ votes.

To ensure statistical reliability and account for variations in model training, we report the mean and standard error of $P$ averaged over $R=10$ independent repetitions of this entire voting protocol. Each repetition $r$ involves a fresh ensemble of $10$ trained models, yielding a corresponding accuracy $P_{r}$  calculated via Eq.~\ref{P}. The final reported accuracy $\bar{P}$ is the arithmetic mean across all repetitions:
\begin{equation}
\bar{P} = \frac{1}{R} \sum_{r=1}^{R} P_r.
\end{equation}
To quantify the uncertainty stemming from the training process, we calculate the standard error (SE) of the mean as:
\begin{equation}
\text{SE} = \sqrt{\frac{\sum_{r=1}^{R} (P_r - \bar{P})^2}{R(R-1)}}.
\end{equation} 
This standard error is represented by the error bars in relevant figures of the main text.

\section{Multilayer perceptron baseline}
To highlight the advantages of the attention mechanism, we compare our primary model against a Multilayer Perceptron (MLP) baseline. While our proposed architecture processes trajectories in sets of $N=25$, the MLP baseline processes each trajectory individually with $N=1$. This comparison allows us to evaluate the necessity of collective trajectory processing and the specific impact of the attention module on classification performance.

The input to the MLP consists of flattened measurement records from the three channels. For a single trajectory $i$, the $X$ channel has a tensor shape of $[T,L]$, the $ZZ$ channel has a shape of $[T/2,L-1]$, and the $ZXZ$ channel has a shape of $[T/3,L-2]$. These tensors are flattened and concatenated into a single input vector $V_{i}$ with a total dimension of $D = TL + \frac{T}{2}(L-1) + \frac{T}{3}(L-2)$. This composite vector represents the complete set of measurement outcomes for one experimental realization.

The MLP architecture follows a sequential structure consisting of three linear layers. The input vector  $V_{i}$ is first projected into a hidden representation of fixed dimension $128$, then into a second hidden layer of variable dimension $h$, as follows: 
\begin{align}
V_{i}^{out} = \text{ReLU}\left[\left(  \text{ReLU}\left(V_{i}W_{1} + b_{1}\right) \right)W_{2}+b_{2}\right]
\end{align}
where $V_{i}^{out} \in \mathbb{R}^{h}$ is the output vector for the $i$-th trajectory, $W_{1}$ and $W_{2}$ are the weight matrices of the first and second hidden layers, and $b_{1}$ and $b_{2}$ are their corresponding bias vectors. The resulting representation $V_{i}^{out}$ is then passed through a final linear layer, which maps it to a three-dimensional output and produces a probability distribution via the softmax activation: 
\begin{equation}
y(i) = \text{Softmax}\left( V_{i}^{out} W_{out} + b_{out} \right).
\end{equation}
where $W_{out}$ and $b_{out}$ are the weight matrix and bias vector of the output layer. The final ensemble prediction $y(M)$ for a given parameter point is obtained by averaging the predictions across all $M$ trajectories as: 
\begin{equation}
y(M) = \frac{1}{M}\sum_{i=1}^{M}y(i)
\end{equation}

\begin{figure}[ht]
\centering
\includegraphics[width=0.45\textwidth, keepaspectratio]{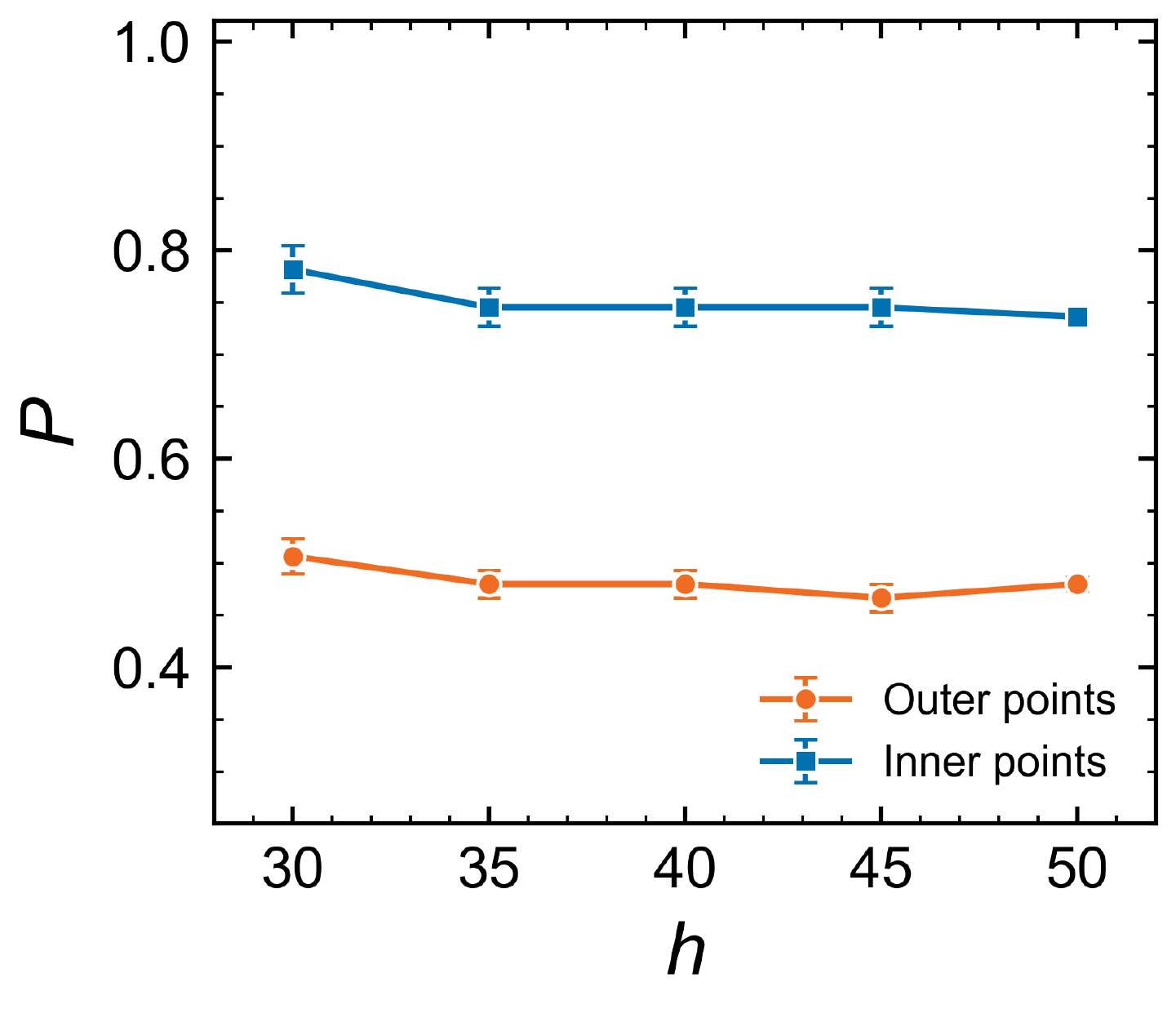}
\caption{Classification accuracy $P$ as a function of hidden layer dimension $h$. Results are shown for  $M=10000$ trajectories per test point. Blue and orange markers correspond to inner and outer test points, respectively. Error bars indicate the standard error across $5$ independent runs.}
\label{fig:MLP}
\end{figure}

The results of this comparison are illustrated in Fig.~\ref{fig:MLP}. We observe a significant degradation in classification accuracy $P$ for the MLP model compared to our proposed attention-based network. For inner test points, the MLP accuracy reaches only approximately $0.8$, whereas our primary model achieves near-perfect classification. The performance gap is even more pronounced for outer test points located near phase boundaries, where the MLP accuracy drops to roughly $0.5$, compared to over $0.8$ for our proposed model. This behavior indicates that observing single trajectories in isolation is insufficient for reliable phase identification. The superior performance of the attention-based architecture confirms that integrating information across multiple trajectories is essential for capturing the underlying entanglement structures, particularly in regions where measurement outcomes are highly stochastic.

